
\overfullrule=0mm
\magnification=1200
\noindent{\bf Moments of the characteristic polynomial in the three 
ensembles of random matrices}
\smallskip

\centerline{Madan Lal Mehta\footnote{*}{Member of CNRS, France.
E-mail address: mehta@spht.saclay.cea.fr} and
Jean-Marie Normand\footnote{**}{E-mail address: norjm@spht.saclay.cea.fr}}

\centerline{CEA/Saclay, Service de Physique Th\'eorique, 
F-91191 Gif-sur-Yvette Cedex, France}
\vskip 1cm

\noindent{\bf Abstract.} Moments of the characteristic polynomial of 
a random matrix taken from any of the three ensembles, orthogonal, 
unitary or symplectic, are given either as a determinant or a 
pfaffian or as a sum of determinants.
For gaussian ensembles comparing the two expressions of the same moment
one gets two remarkable identities, one between an $n\times n$ determinant
and an $m\times m$ determinant and another between the pfaffian
of a $2n\times 2n$ anti-symmetric matrix and a sum of $m\times m$ 
determinants.
\vskip 1cm

\noindent{\bf 1. Introduction}
\bigskip

\noindent Three  ensembles of random matrices $A$ have been extensively 
investigated,
namely when $A$ is real symmetric, complex hermitian or quaternion
self-dual. Recently the behaviour of their characteristic polynomial 
$\det(xI-A)$ has been of interest and we study here its integer moments.
The same method can be adapted to compute the crossed averages
$\langle \prod_{j=1}^m \det(x_jI-A)\rangle$.
\smallskip

Consider a non-negative function $w(x)$ with all its moments finite,
$$\int x^m w(x) dx <\infty \hskip 5mm m=0, 1, ...\ . \eqno (1.1)$$
With this weight function $w(x)$ let us define three scalar products, one
symmetric and two anti-symmetric, as [1] 
$$\eqalignno{ \langle f,g\rangle_2 & :=  \int f(x)
g(x) w(x) dx = \langle g,f\rangle_2 & (1.2) \cr
\langle f,g\rangle_4 & := \int [f(x) g'(x)-f'(x)
g(x)] w(x) dx = -\langle g,f\rangle_4 & (1.3) \cr
\langle f,g\rangle_1 & := \int \int f(x) g(y){\rm
sign}(y-x) \sqrt{w(x) w(y)} dx dy = -\langle g,f\rangle_1 & (1.4) \cr}$$
and introduce polynomials $C_n$, $Q_n$ and $R_n$ of degree $n$, satisfying
the orthogonality relations
$$\eqalignno{ \langle C_n,C_m\rangle_2 & = c_n \delta_{n,m}  & (1.5) \cr
\langle Q_{2n}, Q_{2m}\rangle_4 & = \langle Q_{2n+1},Q_{2m+1}\rangle_4 = 0 
\hskip 5mm
\langle Q_{2n}, Q_{2m+1}\rangle_4 = q_n \delta_{n,m} & (1.6) \cr
\langle R_{2n}, R_{2m}\rangle_1 & = \langle R_{2n+1},R_{2m+1}
\rangle_1 = 0 
\hskip 5mm
\langle R_{2n}, R_{2m+1}\rangle_1 = r_n \delta_{n,m}. & (1.7) \cr} $$
We will
take these polynomials to be monic, i.e. the coefficient of the highest power
will be taken to be one. 
The above conditions determine completely the $C_n(x)$,
$Q_{2n}(x)$ and $R_{2n}(x)$, while $Q_{2n+1}(x)$ can be replaced by
$Q_{2n+1}(x)+aQ_{2n}(x)$ with an arbitrary constant $a$. Similarly,
$R_{2n+1}(x)$ can be replaced by $R_{2n+1}(x)+aR_{2n}(x)$ with
arbitrary $a$. We will choose these constants so that the coefficient of
$x^{2n}$ in $Q_{2n+1}(x)$ and in $R_{2n+1}(x)$ is zero, thus fixing them
also completely.
\smallskip

The subscript $\beta=1$, 2 or 4 is used to remind that it was convenient 
to use these polynomials to express the correlation functions of the 
eigenvalues of real symmetric, complex hermitian and quaternion self-dual 
random matrices respectively [2]. The polynomials $C_j(x)$ are said to be
orthogonal while the polynomials $Q_j(x)$ and $R_j(x)$ are said to be
skew-orthogonal.
\medskip

In what follows $n$ and $m$ are non-negative integers. It can be checked
that all the given formulas remain valid on replacing a non-existing
sum by 0, a non-existing product, integral or determinant by 1 and
forgetting non-existing rows or columns in a matrix. 
\medskip

It has long been known that the orthogonal polynomial $C_n(x)$ can be
expressed as a multiple integral [3]
$$C_n(x) \propto \int ... \int \Delta^2_n({\bf y}) 
\prod_{j=1}^n[(x-y_j)w(y_j)dy_j] \eqno (1.8)$$
where 
$$ {\bf y} := \{y_1, ...,y_n\} \eqno (1.9)$$
and
$$ \Delta_n({\bf y}) :=  \prod_{1\le j<k\le n}(y_k-y_j). \eqno (1.10) $$
Similar multiple integral expressions for the skew-orthgonal
polynomials $Q_n(x)$ and $R_n(x)$ were recently discovered by B. Eynard [4]
$$\eqalignno{                                                     
Q_{2n}(x) & \propto \int ... \int \Delta^4_n({\bf y})
\prod_{j=1}^n [(x-y_j)^2 w(y_j)dy_j] & (1.11)  \cr
Q_{2n+1}(x) & \propto \int ... \int \Delta_n^4({\bf y})  
\left(x+2\sum_{j=1}^n y_j\right)\prod_{j=1}^n [(x-y_j)^2 w(y_j)dy_j] 
& (1.12) \cr
R_{2n}(x) & \propto \int ... \int \vert \Delta_{2n}({\bf y}) \vert
\prod_{j=1}^{2n} [(x-y_j)\sqrt{w(y_j)}dy_j] & (1.13)  \cr
R_{2n+1}(x) & \propto \int ... \int \vert \Delta_{2n}({\bf y}) \vert
\left(x+\sum_{j=1}^n y_j\right) \prod_{j=1}^{2n} [(x-y_j)\sqrt{w(y_j)}
dy_j]. & (1.14) \cr} $$
It is also known [5] that
$$I_2(n,m;x):= \int ...\int \Delta_n^2({\bf y})\prod_{j=1}^n
\left[(x-y_j)^mw(y_j) dy_j\right] \eqno (1.15)$$
can be expressed as an $m\times m$ determinant. We report here that 
similar integrals  
$$I_\beta(n,m;x):= \int ...\int \vert\Delta_n({\bf y})\vert^\beta 
\prod_{j=1}^n\left\{(x-y_j)^m[w(y_j)]^{1/(1+\delta_{\beta,1})} dy_j\right\}
\eqno (1.16)$$
for $\beta=1$ or 4, can as well be expressed as a sum of $m\times m$
determinants.
They are the $m$-th moments of the characteristic polynomial of an 
$n\times n$ random matrix $A$ 
$$\langle \det (xI-A)^m\rangle = {I_\beta(n,m;x)\over I_\beta(n,0;x)}
\eqno (1.17)$$
the parameter $\beta$ taking the values 1, 2 or 4 according as $A$ is 
real symmetric, complex hermitian or quaternion self-dual and the 
probability density of the eigenvalues ${\bf y}$ being 
$\vert\Delta_n({\bf y})\vert^\beta
\prod_{j=1}^n [w(y_j)]^{1/(1+\delta_{\beta,1})}$.
The special case $w(y)=e^{-ay^2}$ arises when the probability density
of the random matrix is invariant under any change of basis and the
algebraically independent parameters specifying the matrix elements
are also statistically independent. 

Instead of the real symmetric, complex hermitian and quaternion self-dual
matrices some authors considered the corresponding ensembles of unitary
matrices [6]. The $m$-th moment of the characteristic polynomial in
$x=\exp(i\alpha)$
$$ \int_0^{2\pi}... \int_0^{2\pi}\prod_{1\le j<k\le n}\vert e^{i\theta_j}
-e^{i\theta_k}\vert^{\beta} \prod_{j=1}^n \left[\vert e^{i\alpha}
-e^{i\theta_j}\vert^m d\theta_j\right] \eqno (1.18)$$
is a constant independent of $\alpha$ which was evaluated using Selberg's
integral [6]. In the limit of large $n$ with fixed $m$, this constant has
a close relation with the moments of the absolute values of the Riemann
zeta function $\zeta(z)$ or of the L-functions on the critical line
${\rm Re}\ z=1/2$. Similarly,
the moments of the derivative with respect to $x$ of the characteristic
polynomial are related to the moments of the absolute values of the
derivative $\zeta'(z)$ on the critical line [6].

In case of the real symmetric, complex hermitian and quaternion self-dual
matrices, which are the only ones we consider here, the $m$-th 
moment of the characteristic polynomial is a polynomial of order $mn$,
and we do not know whether their zeros are related with  the Riemann zeta
function.
\vskip  8mm                                      

\noindent{\bf 2. Calculation of ${\bf I_\beta (n,m;x)}$}
\bigskip

\noindent The case $\beta=2$ is the simplest involving the better known
determinants, while the other two cases $\beta=1$ and 4 are similar
involving the less familiar pfaffians.  

\noindent The $I_\beta(n,m;x)$ can be expressed in two different forms: 

\noindent (i) a determinant or a pfaffian of a matrix of size which 
depends on $n$; i.e.

\noindent - an $n\times n$ determinant in case $\beta=2$;

\noindent - a pfaffian of a $2n\times 2n$ anti-symmetric matrix in 
case $\beta=4$;

\noindent - a pfaffian of a $n\times n$ (resp. $(n+1)\times (n+1)$) 
anti-symmetric matrix in case $\beta=1$ for $n$ even (resp. odd);

\noindent (ii) a sum of determinants of matrices of size which depends 
on $m$; i.e. 

\noindent - an $m\times m$ determinant in case $\beta=2$;

\noindent - a sum of $m\times m$ determinants in case $\beta=4$ for $m$ 
even;
                                                                   
\noindent - a sum of $m\times m$ (resp. $(m+1)\times (m+1)$) determinants
in case $\beta=1$ for $n$ even (resp. odd). 

The usefulness of the expressions above, depending on the form of the
weight function $w(x)$, 
may be limited. When the weight is gaussian, two of the forms supplying
alternate expressions for $I_2(n,m;x)$ as $n\times n$ and $m\times m$
determinants lead us to an interesting known identity. Similarly,
the two expressions of $I_4(n,m;x)$ lead us to an identity between
a pfaffian of an $2n\times 2n$ anti-symmetric matrix and a sum of
$m\times m$ determinants.
\vskip 8mm

\noindent{\bf 2.1 ${\bf I_\beta(n,m;x)}$ as a determinant or a pfaffian of
a matrix of size depending on ${\bf n}$}
\bigskip

\noindent Choose $P_j(x)$ and $\pi_j(x)$ any monic polynomials of degree $j$.
\bigskip

\noindent{\bf 2.1.1 The case ${\bf \beta=2}$}
\bigskip

\noindent Write $\Delta_n({\bf y})$ as an $n\times n$ Vandermonde determinant 
$$\eqalignno{
\Delta_n({\bf y}) & = \det\left[y_j^{k-1}\right]_{j,k=1,...,n} 
 = \det\left[P_{k-1}(y_j)\right]_{j,k=1,...,n} & \cr 
& = \det\left[\pi_{k-1}(y_j)\right]_{j,k=1,...,n}\ .
& (2.1.1) \cr}$$
Then 
$$\eqalignno{
\int...\int \Delta_n^2({\bf y}) \prod_{j=1}^n \left[f(y_j)dy_j\right]
& = \int...\int \det\left[P_{k-1}(y_j)\right] 
\det\left[\pi_{k-1}(y_j)\right]\prod_{j=1}^n \left[f(y_j)dy_j\right] & \cr
& = n! \int...\int \det\left[\pi_{k-1}(y_j)\right]
\prod_{j=1}^n \left[P_{j-1}(y_j) f(y_j)dy_j\right] & \cr
& = n! \int...\int \det\left[P_{j-1}(y_j)\pi_{k-1}(y_j)f(y_j)\right]
\prod_{j=1}^n dy_j & \cr
& = n!\ \det\left[\phi_{2;j,k}\right]_{j,k=0,...,n-1} & (2.1.2) \cr}$$
where                                                              
$$ \phi_{2;j,k} := \int P_j(y)\pi_k(y) f(y) dy. \eqno (2.1.3)$$
In the first three lines of the above equation the indices $j$, $k$ 
take the values from 1 to $n$. In the second line we have replaced 
the first determinant by a single term which is allowed by the 
symmetry of the integrand in the $y_j$; while the last line is obtained 
by integrating over the $y_j$ each of them occuring only in one column.

Setting $f(y)=(x-y)^m w(y)$, we get a first form of $I_2(n,m;x)$ in terms
of an $n\times n$ determinant 
$$I_2(n,m;x) = n!\ \det\left[\phi_{2;j,k}(x)\right]_{j,k=0,...,n-1}
\eqno (2.1.4)$$
where
$$ \phi_{2;j,k}(x) := \int P_j(y)\pi_k(y) (x-y)^m w(y) dy. \eqno (2.1.5)$$
\smallskip

\noindent{\bf 2.1.2 The case ${\bf \beta = 4}$}
\bigskip

\noindent Write $\Delta_n^4({\bf y})$ as a
$2n\times 2n$ determinant [7]   
$$\eqalignno{
\Delta^4_n({\bf y})
& = \det\left[y_j^k, \ ky_j^{k-1}\right]_{j=1,...,n;\ k=0,...,2n-1} &  \cr 
& = \det\left[P_k(y_j), \ P'_k(y_j)\right]_{j=1,...,n;\ k=0,...,2n-1}. 
& (2.1.6) \cr}$$
Each variable occurs in two columns, a column of monic polynomials and
a column of its derivatives. Expand the determinant and integrate to see
that [8]
$$\eqalignno{
\int...\int \Delta^4_n({\bf y}) \prod_{j=1}^n f(y_j)dy_j
& = \sum \pm \phi_{4;j_1,j_2}\phi_{4;j_3,j_4}...
\phi_{4;j_{2n-1},j_{2n}} & (2.1.7) \cr 
& = n!\ {\rm pf}\left[\phi_{4;j,k}\right]_{j,k=0,...,2n-1} & (2.1.8) \cr}$$
where
$$\phi_{4;j,k} := \int \left[P_j(y)P'_k(y)-P'_j(y)P_k(y)\right]f(y)dy
= - \phi_{4;k,j}. \eqno (2.1.9)$$
In equation (2.1.7) the sum is over all permutations
$\left( \matrix {0&1&...&2n-1 \cr j_1&j_2&...&j_{2n} \cr} \right)$ of the
$2n$ indices $(0,1,...,2n-1)$ with $j_1<j_2$, ..., $j_{2n-1}<j_{2n}$,
the sign being plus or minus according as
this permutation is even or odd. In equation (2.1.8) {\it pf}
means the pfaffian.

Setting $f(y) = (x-y)^m w(y)$ as before, we get a first form of $I_4(n,m;x)$
in terms of the pfaffian of a $2n\times 2n$ anti-symmetric matrix 
$$ I_4(n,m;x) = n!\ {\rm pf}\left[\phi_{4;j,k}(x)\right]_{j,k=0,...,2n-1}
\eqno (2.1.10)$$
where
$$\phi_{4;j,k}(x) := \int \left[P_j(y)P'_k(y)-P'_j(y)P_k(y)\right]
(x-y)^m w(y) dy. \eqno (2.1.11)$$
\bigskip

\noindent{\bf 2.1.3 The case ${\bf \beta=1}$}
\bigskip
\noindent The difficulty of the absolute
value of $\Delta_n({\bf y})$ can be overcome by ordering the variables.
Writing
$$\Delta_n({\bf y})\prod_{j=1}^n f(y_j)
 = \det \left[P_{k-1}(y_j)f(y_j) \right]_{j,k=1,...,n} \eqno (2.1.12)$$
with $f(y)=(x-y)^m\sqrt{w(y)}$ and using a result of de Bruijn [9],
one gets a first form of $I_1(n,m;x)$ in terms of the pfaffian of
an $n\times n$ (resp. $(n+1)\times (n+1)$) anti-symmetric matrix for
$n$ even (resp. odd) 
$$I_1(n,m;x) = \left\{\matrix {n!\ {\rm pf}\left[ \phi_{1;j,k}(x)
\right]_{j,k=0,...,n-1} \hfill & n\ {\rm even}  \cr 
n!\ {\rm pf}\left[
\matrix{\phi_{1;j,k}(x) & \alpha_j(x) \cr -\alpha_k(x) & 0 \cr} 
\right]_{j,k=0,...,n-1} & n\ {\rm odd} \cr} \right. \eqno (2.1.13)$$
where
$$\phi_{1;j,k}(x) := \int\int P_j(z)P_k(y) (x-z)^m (x-y)^m
\sqrt{w(z)w(y)} {\rm sign}(y-z) dz dy \eqno (2.1.14) $$
and
$$ \alpha_j(x) := \int P_j(y) (x-y)^m \sqrt{w(y)} dy. \eqno (2.1.15)$$

Another way to get the same result is [8] to order the variables, integrate
over the alternate ones, remove the ordering over the remaining alternate
variables and observe that the result is given by equations
(2.1.13)-(2.1.15).
\vskip 8mm

\noindent{\bf 2.1.4 An alternative of the expressions above for ${\bf 
\beta=1}$, 2 and 4}
\bigskip

\noindent If one replaces the monic polynomials $P_j(y)$ and $\pi_j(y)$ by
$P_j(y-x)$ and $\pi_j(y-x)$, equations (2.1.4), (2.1.10) and (2.1.13) 
become  
$$\eqalignno{
I_2(n,m;x) & = (-1)^{mn}n!\ \det\left[\varphi_{2;j,k}(x)
\right]_{j,k=0,...,n-1} & (2.1.16) \cr 
I_4(n,m;x) & = (-1)^{mn} n!\ {\rm pf}\left[\varphi_{4;j,k}(x)
\right]_{j,k=0,...,2n-1} & (2.1.17) \cr 
I_1(n,m;x) & = \left\{\matrix{n!\ {\rm pf}\left[ \varphi_{1;j,k}(x)
\right]_{j,k=0,...,n-1} \hfill & n\ {\rm even}  \cr 
(-1)^m n!\ {\rm pf}\left[
\matrix{\varphi_{1;j,k}(x) & \theta_j(x) \cr -\theta_k(x) & 0 \cr} 
\right]_{j,k=0,...,n-1} & n\ {\rm odd} \cr} \right. & (2.1.18) \cr}$$
where 
$$\eqalignno{
\varphi_{2;j,k}(x) & := \int P_j(y-x)\pi_k(y-x) (y-x)^m w(y) dy & 
(2.1.19) \cr
\varphi_{4;j,k}(x) & := \int \left[P_j(y-x)P'_k(y-x)-P'_j(y-x)P_k(y-x)\right]
(y-x)^m w(y) dy & (2.1.20) \cr
\varphi_{1;j,k}(x) & := \int\int P_j(z-x) P_k(y-x) [(z-x)(y-x)]^m
\sqrt{w(z) w(y)}{\rm sign}(y-z) dz dy & \cr & & (2.1.21) \cr
\theta_j(x) & := \int P_j(y-x) (y-x)^m \sqrt{w(y)} dy. & (2.1.22) \cr}$$
\vskip 8mm 

\noindent{\bf 2.2\ ${\bf I_\beta(n,m;x)}$ as determinants of size
depending on $m$}
\bigskip

\noindent{\bf 2.2.1 The case ${\bf \beta=2}$}
\bigskip

\noindent For $I_2(n,m;x)$, equation (1.15), let us write the integrand 
as the product of two determinants:  
$\Delta_n({\bf y})$  given in equation (2.1.1) and 
$$\eqalignno{                                    
\Delta_n({\bf y})\prod_{j=1}^n(x-y_j)^m 
& = b\ \det\left[y_j^{k-1}, \left({d\over dx}\right)^l x^{k-1}\right]
_{\matrix{ \scriptstyle
j=1,...,n\hfill \cr \scriptstyle k=1,...,n+m\hfill\cr 
\scriptstyle l=0,...,m-1 \hfill \cr}} & \cr
& = b\ \det\left[P_{k-1}(y_j), P^{(l)}_{k-1}(x)\right]_{\matrix{
\scriptstyle j=1,...,n\hfill\cr \scriptstyle k=1,...,n+m\hfill\cr 
\scriptstyle l=0,...,m-1 \hfill \cr}} & (2.2.1)
\cr}$$
where
$$b = \left(\prod_{l=0}^{m-1}l!\right)^{-1} \eqno (2.2.2)$$
$P_k(x)$ is any monic polynomial of degree $k$ and $P_k^{(l)}(x)$ is its
$l$-th derivative. Expression (2.2.1) is an $(m+n)\times (m+n)$ determinant
the last $m$ columns of which are $P_k(x)$ and its successive derivatives 
[7]. As the integrand in equation (1.15) is symmetric in the $y_j$, we 
can replace the  first $n\times n$ determinant by a single term and multiply 
the result by $n!$ 
$$\eqalignno{
I_2(n,m;x) & = b\ n! \int ...\int
\det \left[ P_{k-1}(y_j), P_{k-1}^{(l)}(x)
\right]_{\matrix{\scriptstyle j=1,...,n \hfill\cr 
\scriptstyle k=1,...,n+m \hfill\cr \scriptstyle l=0,...,m-1
\hfill \cr}} 
\prod_{j=1}^n \left[P_{j-1}(y_j)w(y_j) dy_j\right] & \cr 
& = b\  n!\int ...\int\det \left[P_{j-1}(y_j)P_{k-1}(y_j)w(y_j), 
P_{k-1}^{(l)}(x)\right]_{\matrix{\scriptstyle j=1,...,n\hfill \cr 
\scriptstyle k=1,...,n+m\hfill \cr \scriptstyle l=0,...,m-1
\hfill \cr}} \prod_{j=1}^n dy_j. & \cr & & (2.2.3) \cr}$$ 
As each variable $y_j$ occurs only in one column, we can integrate over 
them independently 
$$I_2(n,m;x) = b\ n! \det\left[\int P_{j-1}(y) P_{k-1}(y) w(y) dy,
P_{k-1}^{(l)}(x)
\right]_{\matrix{\scriptstyle j=1,...,n \hfill\cr 
\scriptstyle k=1,...,n+m\hfill \cr \scriptstyle l=0,...,m-1 \cr}}. 
\eqno (2.2.4) $$
If we choose the polynomials $P_j(x)$ to be the orthogonal polynomials
$C_j(x)$, equation  (1.5), then one gets a second form for $I_2(n,m;x)$ 
$$I_2(n,m;x) = b\ n!\ c_0...c_{n-1} \det                                 
\left[C_{n+k}^{(l)}(x)
\right]_{k,l=0,...,m-1} \eqno (2.2.5)$$
an $m\times m$ determinant whose first column consists of $C_{n+k}(x)$, 
$k=0$, 1, ..., $m-1$, and the other $m-1$ columns are the successive 
derivatives of the first column. Therefore  from equation (1.17) 
$$\langle \det(xI-A)^m\rangle = b\ \det\left[C_{n+k}^{(l)}(x)
\right]_{k,l=0,...,m-1}. \eqno (2.2.6)$$
This result appears in reference [5] as equation (15) apart from a sign
which seems to be wrong.
\vskip 8mm

\noindent{\bf 2.2.2 The case ${\bf \beta=4}$}
\bigskip

\noindent This method applies only for even moments. 
For $I_4(n,2m;x)$, equation (1.16), one can write the integrand
as a single determinant [7]  
$$\eqalignno{
\Delta^4_n({\bf y})\prod_{j=1}^n(x-y)^{2m} & = b\ 
\det\left[y_j^k, ky_j^{k-1}, \left({d\over dx}\right)^l x^k\right]
_{\matrix{\scriptstyle j=1,...,n\hfill \cr 
\scriptstyle k=0,...,n+m-1\hfill \cr \scriptstyle l=0,...,m-1
\hfill \cr}} & \cr
& = b\ \det\left[P_k(y_j), P'_k(y_j), P^{(l)}_k(x)\right]
_{\matrix{\scriptstyle j=1,...,n\hfill \cr 
\scriptstyle k=0,...,n+m-1\hfill \cr \scriptstyle l=0,...,m-1
\hfill \cr}} & (2.2.7) \cr}$$
where $b$ is given by equation  (2.2.2). 
Each variable $y_j$ occurs in two columns. Expanding the determinant
and integrating one sees that the result is a sum of products of the form 
$$ \pm b\ a_{s_1,s_2}... a_{s_{2n-1},s_{2n}}\det 
\left[P_{j_k}^{(l-1)}(x) \right]_{k,l=1,...,m} \eqno (2.2.8)$$
with  
$$a_{j,k} := \int \left[P_j(y)P'_k(y)-P'_j(y)P_k(y)\right]w(y) dy
= \langle P_j,P_k\rangle_4 \eqno (2.2.9)$$
the indices $s_1$, ..., $s_{2n}$, $j_1$, ..., $j_m$ are all 
distinct, chosen from 0, 1, ..., $2n+m-1$ and the sign is plus or minus
according as the permutation 
$$\left( \matrix{0&1&...&2n-1&2n&...&2n+m-1\cr 
s_1&s_2&...&s_{2n}&j_1&...&j_m\cr} \right)  \eqno (2.2.10)$$
is even or odd. If we choose the polynomials $P_j(x)$ to be the
skew-orthogonal polynomials $Q_j(x)$, equation  (1.6), then all
the $a_{s,j}$ except $a_{2s,2s+1}=q_s$ will be zero and we get
a second form of $I_4(n,2m;x)$ in terms of a  sum of $m\times m$
determinants   
$$I_4(n,2m;x) = b\ n!\sum_{(s)} q_{s_1}...q_{s_n} \det \left[
Q^{(l-1)}_{j_k}(x) \right]_{k,l=1,...,m}  \eqno (2.2.11)$$
where the sum is over all choices of $s_1<...<s_n$ such that $2s_1$,
$2s_1+1$, ..., $2s_n$, $2s_n+1$, $j_1$, ..., $j_m$ are all the
indices  from 0 to $2n+m-1$ and moreover $j_1<...<j_m$.
\vskip 8mm     

\noindent{\bf 2.2.3 The case ${\bf \beta=1}$}
\bigskip

\noindent For $I_1(n,m;x)$, equation (1.16), the absolute value sign 
of $\Delta_n({\bf y})$ is the main difficulty. Ordering the variables 
and using the same method as in section 2.2.2 above [7], one has
$$\eqalignno{                        
I_1(n,m;x) & = \int ...\int \vert \Delta_n({\bf y})\vert \prod_{j=1}^n 
\left[(x-y_j)^m \sqrt{w(y_j)} dy_j\right] & \cr 
& = n! \int...\int_{y_1\le ...\le y_n} 
\Delta_n({\bf y}) \prod_{j=1}^n 
\left[(x-y_j)^m \sqrt{w(y_j)} dy_j\right] & \cr 
& = b\ n! \int...\int_{y_1\le ...\le y_n}  
\det\left[P_k(y_j), P_k^{(l)}(x) \right]_{
\matrix{ \scriptstyle j=1,...,n\hfill \cr  
\scriptstyle k=0,...,n+m-1\hfill \cr
\scriptstyle l=0,...,m-1\hfill \cr }} & \cr
& \hskip 1cm .\prod_{j=1}^n \left[\sqrt{w(y_j)} dy_j\right] & (2.2.12) \cr }$$
with $b$ as in equation  (2.2.2).
Integrating successively over the alternate variables $y_1$, $y_3$,
$y_5$, ... and then removing the restriction over the remaining
variables $y_2$, $y_4$, ... as indicated at the end of section 2.1.3 above,
one gets
for even $n$ 
$$ \eqalignno{I_1(n,m;x) & = b\ {n!\over [n/2]!} \int ... \int 
\det\left[G_k(y_{2j}), P_k(y_{2j}), P^{(l)}_k(x)
\right]_{\matrix{\scriptstyle j=1,...,[n/2] \hfill \cr 
\scriptstyle k=0,...,n+m-1 \hfill \cr
\scriptstyle l=0,...,m-1 \hfill \cr}} & \cr
& \hskip 1cm .\prod_{j=1}^{[n/2]} \left[\sqrt{w(y_{2j})} dy_{2j}\right] 
& (2.2.13) \cr}$$
with 
$$G_k(y_2) := \int^{y_2} P_k(y_1)\sqrt{w(y_1)} dy_1. \eqno (2.2.14)$$
In case $n$ is odd, one more column of the numbers $g_k:=\int P_k(y)
\sqrt {w(y)}dy$ appears in the
determinant just after the column of $P_k(y_{n-1})$.

The present situation is exactly as in the case $\beta=4$ with each
variable occuring in two columns. Expanding the determinant one sees that 
the result contains the expressions 
$$\eqalignno{
\int \left[G_j(y)P_k(y)-G_k(y)P_j(y)\right]\sqrt{w(y)}dy
& = \int\int P_j(x)P_k(y){\rm sign}(y-x) \sqrt{w(x)w(y)} dx dy & \cr
& = \langle P_j,P_k\rangle_1. & (2.2.15) \cr}$$
If we choose $P_j(x)= R_j(x)$, equation  (1.7), then writing the
result separately for even and odd $n$ for clarity, one has 
$$I_1(2n,m;x) = b\ (2n)! \sum_{(s)}r_{s_1}...r_{s_n}
\det\left[ R^{(l-1)}_{j_k}(x)\right]_{k,l=1,...,m} \eqno (2.2.16)$$
where the sum is taken over all choices of $s_1<...<s_n$ such that
$2s_1$, $2s_1+1$, ..., $2s_n$, $2s_n+1$, $j_1$, ..., $j_m$
are all the indices from 0 to $2n+m-1$ and moreover $j_1<...<j_m$ 
and 
$$I_1(2n+1,m;x) = b\ (2n+1)! \sum_{(s)}r_{s_1}...r_{s_n}
\det\left[g_{j_k}, R^{(l-1)}_{j_k}(x)\right]_{k=1,...,m+1;
\ l=1,...,m} \eqno (2.2.17)$$
where now the sum is taken over all choices of $s_1<...<s_n$ such that
$2s_1$, $2s_1+1$, ..., $2s_n$, $2s_n+1$, $j_1$, ..., $j_m$, $j_{m+1}$ 
are all the indices from 0 to $2n+m$ and moreover $j_1<...<j_{m+1}$.
\smallskip

Note that equations (1.8), (1.11) and (1.13) are particular cases ($m=1$) 
of equations (2.2.5), (2.2.11) and (2.2.16) respectively.
If we shift the index of the last row in the right hand side of
equation (2.2.7) for $m=1$ from $n$ to $n+1$, i.e. replace the last row 
$$[P_{n}(y_j),P'_{n}(y_j),P_n(x)]$$
by the row 
$$[P_{n+1}(y_j),P'_{n+1}(y_j),P_{n+1}(x)]$$
we get the integrand of equation (1.12). Following the procedure which 
leads to equation (2.2.11) we get equation (1.12). Similarly, if we shift
the index of the last row from $n$ to $n+1$ in the right hand side of
equation (2.2.12) for $m=1$ and follow the procedure leading to equation 
(2.2.16)  we get equation (1.14).
\vskip 8mm

\noindent{\bf 3. Special case of the gaussian weight}
\bigskip

\noindent These formulas for $I_\beta(n,m;x)$ will have little use if 
one does not know the polynomials $C_n(x)$, $Q_n(x)$ or $R_n(x)$. 
Fortunately one knows them for almost all classical weights. As 
an example we give them here for the gaussian weight 
$w(x)= e^{-x^2}$ over $[-\infty,\infty]$ in terms of Hermite 
polynomials $H_n(x):=e^{x^2}(-d/dx)^ne^{-x^2}$. Their verification is 
straightforward. One has 
$$ \eqalignno{
C_n(x) & = 2^{-n}H_n(x) & (3.1) \cr
c_n & = 2^{-n} n!\sqrt{\pi} & (3.2) \cr
Q_{2n}(x) & = \sum_{j=0}^n 2^{-2j}{n!\over j!}H_{2j}(x) 
\hskip 5mm 
Q_{2n+1}(x) = 2^{-2n-1}H_{2n+1}(x) & (3.3) \cr
q_n & = 2^{-2n} (2n+1)! \sqrt{\pi} & (3.4) \cr 
R_{2n}(x) & = 2^{-2n} H_{2n}(x) 
\hskip 5mm
R_{2n+1}(x) = 2^{-2n}[xH_{2n}(x) - H'_{2n}(x)] & (3.5) \cr
r_n & = 2^{1-2n} (2n)! \sqrt{\pi}. & (3.6) \cr}$$
\smallskip

In this case some results can be given other forms. In particular, 
using the recurrence relation
$$2xH_n(x) = H_{n+1}(x)+H'_n(x) \eqno (3.7) $$
or
$$xC_n(x) = C_{n+1}(x)+{1\over 2}C'_n(x) \eqno (3.8) $$
one can replace the $C_{n+k}^{(l)}$ in the determinant (2.2.5) 
by $C_{n+k+l}(x)$ so that
$$ \eqalignno{ I_2(n,m;x) & = b\ n!\ c_0...c_{n-1}(-2)^{m(m-1)/2}
\det\left[ C_{n+j+k}(x)\right]_{j,k=0,1,...,m-1} & \cr 
& = \pi^{n/2}{2^{-n(n-1)/2} \prod_{j=0}^n j!
\over (-2)^{-m(m-1)/2} \prod_{j=0}^{m-1}j!}
\det\left[ C_{n+j+k}(x)\right]_{j,k=0,1,...,m-1}. & (3.9) \cr} $$
Also for any non-negative integer $j$ [10] 
$$ \int_{-\infty}^\infty (y-x)^j e^{-y^2} dy = \sqrt{\pi}\ i^j\ C_j(ix) 
\eqno (3.10)$$ 
with $i=\sqrt{-1}$. Therefore, choosing $P_j(y-x)=\pi_j(y-x)=(y-x)^j$ in 
equations (2.1.19) and (2.1.20), we get 
$$ \eqalignno{
\varphi_{2;j,k}(x) & = \sqrt{\pi}\ i^{m+j+k}\ C_{m+j+k}(ix) & (3.11) \cr
\varphi_{4;j,k}(x) & = \sqrt{\pi}\ i^{m+j+k-1}\ (k-j) C_{m+j+k-1}(ix) 
& (3.12) \cr}$$
and equations (2.1.16) and (2.1.17) then give 
$$ \eqalignno{ 
I_2(n,m;x) & = (-i)^{mn}\ n!\ \pi^{n/2}(-1)^{n(n-1)/2} 
\det\left[C_{m+j+k}(ix)\right]_{j,k=0, ...,n-1} & (3.13) \cr 
I_4(n,m;x) & = (-i)^{mn}\ n!\ \pi^{n/2} 
{\rm pf}\left[(k-j) C_{m+j+k-1}(ix) \right]_{j,k=0,...,2n-1}\ .
& (3.14) \cr}$$

Equations (3.9) and (3.13) give a relation almost symmetric in $n$ and $m$. 
Writing them again 
$${I_2(n,m;x)\over I_2(m,n;ix)}
= (-i)^{mn}{\pi^{n/2}2^{-n(n-1)/2}\prod_{j=0}^n j!
\over \pi^{m/2}2^{-m(m-1)/2}\prod_{j=0}^m j!}  \eqno(3.15) $$
or equivalently 
$$ {\det\left[C_{n+j+k}(x)\right]_{j,k=0,...,m-1}\over 
\det\left[C_{m+j+k}(ix)\right]_{j,k=0,...,n-1}}
= (-i)^{mn} {(-2)^{n(n-1)/2}\prod_{j=0}^{n-1} j!\over 
(-2)^{m(m-1)/2}\prod_{j=0}^{m-1} j!}. \eqno (3.16)$$
Equations (3.15) and (3.16) appear in [11] as equations (4.43) and (4.44).
\medskip

Equations (2.2.11) and (3.14) give another identity relating the
pfaffian 
$${\rm pf}[(k-j)C_{2m+j+k-1}]_{j,k=0,...,2n-1}$$ 
to a sum of $m\times m$ determinants.
\medskip

Forrester and Witte [11] state that the identity 
$${\int_{-\infty}^\infty ... \int_{-\infty}^\infty 
\vert\Delta_n({\bf y})\vert^{2\gamma}\prod_{j=1}^n\left[(x-y_j)^m
e^{-y_j^2}dy_j\right]
\over \int_{-\infty}^\infty ... \int_{-\infty}^\infty 
\vert\Delta_m({\bf y})\vert^{2/\gamma}\prod_{j=1}^m\left[(ix-y_j)^n
e^{-y_j^2}dy_j\right]} = {\rm const.} \eqno (3.17) $$
is valid  for any $\gamma$. This is not true for $\gamma\ne 1$ as verified
below for $\{n,m\}=\{n,1\},\ \{2,2\}$ and $\{3,2\}$.
For values of $a$ and $\gamma$ such that the integrals  below exist, letting
$$\eqalignno{
\int d\mu({\bf y}) f({\bf y}) & :=
\int_{-\infty}^\infty ... \int_{-\infty}^\infty 
\vert\Delta_n({\bf y})\vert^{2\gamma}\prod_{j=1}^n 
\left(e^{-ay_j^2}dy_j\right) f({\bf y}) & (3.18) \cr  
\langle f({\bf y})\rangle & := \int d\mu({\bf y})f({\bf y}) \div 
\int d\mu({\bf y}) & (3.19) \cr
{\cal I}(\gamma,a,n,m;x) & :=
\int d\mu({\bf y})\prod_{j=1}^n (x-y_j)^m
= {\cal I}(\gamma,a,n,0;x) \langle \prod_{j=1}^n(x-y_j)^m\rangle
& (3.20) \cr }$$
where $\langle \prod_{j=1}^n(x-y_j)^m\rangle $ is a polynomial in
$x$ whose coefficients can be evaluated using the expressions given in
[12].
\medskip

\noindent For any $n$ and $m=1$, one has
$$\eqalignno{
{{\cal I}(\gamma,a,n,1;x)\over {\cal I}(\gamma,a,n,0;x)} & =
\langle \prod_{j=1}^n(x-y_j)\rangle
= \sum_{j=0}^n\left(\matrix {n\cr j \cr}\right)(-1)^j
\langle y_1...y_{j}\rangle x^{n-j} & \cr
& = \sum_{j=0}^{[n/2]}\left(\matrix {n\cr 2j \cr}\right)
\left(-{\gamma\over 2a}\right)^j {(2j)!\over 2^j j!} x^{n-2j} & \cr
& = \left({\gamma\over 4a}\right)^{n/2}H_n
\left(x\sqrt{{a\over\gamma}}\right) & (3.21) \cr}$$
whereas for $n=1$ and any $m$ 
$$\eqalignno{
{{\cal I}(\gamma,a,1,m;x)\over {\cal I}(\gamma,a,1,0;x)}
& = \langle (x-y_1)^m \rangle 
= \sum_{j=0}^m \left(\matrix {m\cr j \cr}\right) (-1)^j
\langle y_1^j \rangle x^{m-j} & \cr 
& = \sum_{j=0}^{[m/2]} \left(\matrix {m\cr 2j \cr}\right) 
{(2j)!\over 2^{2j} j!} a^{-j} x^{m-2j} & \cr
& = (4a)^{-m/2} i^{-m} H_m(ix\sqrt{a}). & (3.22) \cr}$$
Therefore for any $\gamma $, $\gamma'$, $a$ and $n$, one verifies that
$${{\cal I}(\gamma,a,n,1;x)\over {\cal I}(\gamma',a/\gamma,1,n;ix)}
= (-i)^n {{\cal I}(\gamma,a,n,0;x)\over {\cal I}(\gamma',a/\gamma,1,0;ix)}
\eqno (3.23)$$
is a constant independent of $x$.

\noindent For $n=m=2$, one has 
$$\eqalignno{
{{\cal I}(\gamma,a,2,2;x)\over {\cal I}(\gamma,a,2,0;x)}
& =\langle \prod_{j=1}^2(x-y_j)^2\rangle
= x^4+2(\langle y_1^2 \rangle + 2\langle y_1y_2 \rangle )x^2 +
\langle y_1^2 y_2^2 \rangle & \cr
& = x^4+{1\over a}(1-\gamma)x^2+{1\over 4a^2}(1+\gamma+\gamma^2)
& \cr
& = {{\cal I}(1/\gamma,a/\gamma,2,2;ix)\over
{\cal I}(1/\gamma,a/\gamma,2,0;ix)} .  & (3.24) \cr}$$ 

\noindent For $n=3$ and $m=2$, one has
$$\eqalignno{
{{\cal I}(\gamma,a,3,2;x)\over {\cal I}(\gamma,a,3,0;x)}
& =\langle \prod_{j=1}^3(x-y_j)^2\rangle & \cr 
& = x^6 + 3(\langle y_1^2 \rangle + 4\langle y_1y_2 \rangle )x^4 +
3(\langle y_1^2y_2^2 \rangle + 4\langle y_1^2y_2y_3 \rangle )x^2 +  
\langle y_1^2 y_2^2 y_3^2 \rangle & \cr
& = x^6+{3\over 2a}(1-2\gamma)x^4+{3\over 4a^2}(1-\gamma+3\gamma^2)x^2
+{1\over 8a^3}(1+3\gamma+5\gamma^2) & \cr & & (3.25) \cr}$$
whereas for $n=2$ and $m=3$ 
$$\eqalignno{
{{\cal I}(\gamma,a,2,3;x)\over {\cal I}(\gamma,a,2,0;x)}
& =\langle \prod_{j=1}^2(x-y_j)^3\rangle & \cr 
& = x^6 + 3(2\langle y_1^2 \rangle + 3\langle y_1y_2 \rangle )x^4 +
3(2\langle y_1^3y_2 \rangle + 3\langle y_1^2y_2^2 \rangle )x^2 +  
\langle y_1^3 y_2^3 \rangle & \cr
& = x^6+{3\over 2a}(2-\gamma)x^4+{3\over 4a^2}(3-\gamma+\gamma^2)x^2
-{\gamma\over 8a^3}(5+3\gamma+\gamma^2) & \cr & & (3.26) \cr}$$
hence 
$${{\cal I}(\gamma,a,3,2;x)\over {\cal I}(\gamma,a,3,0;x)}
=-{{\cal I}(1/\gamma,a/\gamma,2,3;ix)\over
{\cal I}(1/\gamma,a/\gamma,2,0;ix)}. \eqno (3.27)$$
Selberg's integral also gives the constant [12]
$${\cal I}(\gamma,a,n,0;x)=\pi^{n/2}
2^{-\gamma n(n-1)/2} a^{-\gamma n(n-1)/2-n/2}
\prod_{j=1}^n {\Gamma(1+j\gamma)\over \Gamma(1+\gamma)}. \eqno (3.28)$$
So, what is probably true is
$${{\cal I}(\gamma,a,n,m;x)\over {\cal I}(1/\gamma,a/\gamma,m,n;ix)}
= (-i)^{mn} {{\cal I}(\gamma,a,n,0;x)\over
{\cal I}(1/\gamma,a/\gamma,m,0;ix)}       \eqno (3.29)$$
a constant  depending on $\gamma$, $a$, $n$ and $m$. 
i.e. the gaussians in the numerator and denominator of equation (3.17)
should have different variances. In particular,
$${I_4(n,m;x)\over I_1(m,n;ix)} = 
(-i)^{mn}{\pi^{n/2} 2^{-n^2}\prod_{j=0}^n (2j)!\over
2^{3m/2}\prod_{j=0}^m \Gamma(1+j/2)}.  \eqno (3.30) $$
\vskip 8mm

\noindent{\bf 4. Conclusion}
\bigskip 

\noindent The $m$-th moment of the characteristic polynomial of an
$n\times n$ random matrix is expressed as: an $n\times n$, equations 
(2.1.4 and 2.1.16), or as an $m\times m$, equation (2.2.5), determinant,
for the unitary ensemble ($\beta=2$); as a 
pfaffian of a $2n\times 2n$ anti-symmetric matrix, equations (2.1.10 and
2.1.17),
or a sum of $m\times m$ determinants in case $m$ is even, equation
(2.2.11), for the symplectic ensemble ($\beta=4$); as a 
pfaffian of an $n\times n$ (resp. $(n+1)\times (n+1)$)
anti-symmetric matrix, equations (2.1.13 and 2.1.18), 
or a sum of $m\times m$ (resp. $(m+1)\times (m+1)$) determinants,
equation (2.2.16, resp. 2.2.17), for the
orthogonal ensemble ($\beta=1$) for $n$ even (resp. odd).

\noindent In the gaussian case $w(y)=e^{-y^2}$ this leads to the remarkable
identity almost symmetric in $m$ and $n$, equation (3.15),  
$$ I_2(n,m;x) = {\rm const.}\ I_2(m,n;ix) \eqno (4.1)$$
or, equation (3.16), 
$$ \det\left[H_{m+j+k}(x)\right]_{j,k=0,...,n-1} = {\rm const.}\ 
\det\left[H_{n+j+k}(ix)\right]_{j,k=0,...,m-1} \eqno (4.2)$$
and another identity expressing
${\rm pf}\left[(k-j)H_{2m+j+k}(ix)\right]_{j,k=0,...,2n-1}$,  
equation (3.14), as a sum
of $m\times m$ determinants, equation (2.2.11).
Also probably the remarkable general identity (3.29) and in particular
(3.30) holds. 
\vskip 8mm

\noindent{\bf Acknowlegements}
\bigskip

\noindent One of us (MLM) is thankful to A. Edelman
for asking the question about the moments of the characteristic 
polynomial. We thank P.J. Forrester for sending us the preprint with N.S. 
Witte and indicating the relevant place in his paper with Baker.
In particular, he indicated that in [11] equations (5.30a), (5.1) and (1.3a)
relate the integral ${\cal I}(\beta/2,\beta/2,n,\beta;x)$, with our
notations in equation (3.20), to a generalized Hermite polynomial which
appears in equation (5.31) of [11] as the integral
${\cal I}(2/\beta,1,\beta,n;ix)$. Thus it provides a proof of conjecture
(3.29) in the special case $m=2\gamma=2a=\beta$. 
We are grateful to our colleague
R. Balian for critically reading and commenting the manuscript.
In particular he noted that if in equation (1.4) one replaces
sign$(y-x)$ by sign$(y-x+\epsilon/2)$ and expands in powers of $\epsilon$,
then the three scalar products are just the first three terms in
this expansion
$$ \int \int f(x) g(y){\rm sign}(y-x+\epsilon/2) \sqrt{w(x) w(y)} dx dy
= \langle f,g\rangle_1
+\epsilon\langle f,g\rangle_2 +{\epsilon^2\over 8}\langle f,g\rangle_4+...
\eqno (4.3)$$
Does this have some deeper significance and/or applications?

We are thankful to the referees for bringing to our attention the references
6 and 9.
\vskip 8mm

\noindent{\bf References}
\bigskip 

\item{[1]} See for example, {\it Matrix theory}, M.L. Mehta, Les Editions de
Physique, 91944 Les Ulis Cedex, France (1989) appendix A.14

\item{[2]} G. Mahoux and M.L. Mehta, A method of integration over matrix
variables IV, J. Phys. I France 1 (1991) 1093-1108 

\item{[3]} See for example, {\it Higher transcendental functions} 
A. Erd\'elyi et al. (ed.), Bateman Manuscript project, MacGraw Hill,
New York (1953), vol. 2, \S 10.3 (5); or
G. Szeg\"o, {\it Orthogonal polynomials},
American Mathematical Society, Providence,  R.I. (1939), equation (2.2.10)

\item{[4]} B. Eynard, private communication and
http://xxx.lanl.cond-mat/0012046

\item{[5]} E. Br\'ezin and S. Hikami, Characteristic polynomials of
random matrices, Comm. Math. Phys. 214 (2000)  111-135

\item{[6]} J.P. Keating and N.C. Snaith, Random matrix theory and
$\zeta(1/2+it)$, Comm. Math. Phys. 214 (2000) 57-89

\item{} J.P. Keating and N.C. Snaith, Random matrix theory and
L-functions at $s=1/2$, Comm. Math. Phys. 214 (2000) 91-110

\item{} C.P. Hughes, J.P. Keating and N. O'Connell,
Random matrix theory and the derivative of the Riemann zeta function,
Proc. Royal Soc. London A: Math. 456 (2000) 2611-2627 

\item{[7]} See for example, reference 1, chapter 7.1.2

\item{[8]} See for example, {\it Random matrices}, M.L. Mehta, Academic
Press, New york (1991) chapter 6.5 and appendix A.18

\item{[9]} N.G. de Bruijn, On some multiple integrals involving determinants,
J. Indian Math. Soc. 19 (1955) 133-151; equations (1.2), (4.2)-(4.4) with
$\phi_k(y_j)=P_{k-1}(y_j)(x-y_j)^m\sqrt{w(y_j)}$
                                                     
\item{[10]} See for example, {\it Higher transcendental functions} Bateman
manuscript project, A. Erd\'eyli et al. (eds.), MacGraw Hill, New York (1953)
vol 2, \S 10.13 (30)-(31)

\item{[11]} P.J. Forrester and N.S. Witte, Application of the $\tau$-function
theory of Painlev\'e equations to random matrices: PIV, PII and GUE,
(preprint), equation (4.43) and the paragraph following equation (4.44). 
See also T.H. Baker and P.J. Forrester, The Calogero-Sutherland model 
and generalized classical polynomials, Comm. Math. Phys. 188 (1997) 
175-216, around equations (5.30)-(5.33) 

\item{[12]} See for example, reference 8, chapter 17, equations 
(17.8.5)-(17.8.9), (17.6.7)

\end